\documentclass[11pt,amscd]{amsart}
\usepackage{amsaddr}
\usepackage{graphicx,amssymb,amsmath,amsthm}
\usepackage{amsfonts}

\usepackage[utf8]{inputenc}
\usepackage{epstopdf}
\usepackage[boxed,linesnumbered]{algorithm2e}	
\usepackage{bm}
\usepackage[numbers]{natbib}
\usepackage{color}
\usepackage{listings}
\usepackage{array}

\usepackage[margin=1.25in]{geometry}
\newcolumntype{L}{>{$}l<{$}} 
\newcolumntype{C}{>{$}c<{$}} 

\usepackage{bm,graphicx,textcomp,bbm} 

\usepackage{booktabs}

\usepackage{subcaption}
\usepackage{moreverb}
\usepackage[colorinlistoftodos]{todonotes}

\usepackage{graphicx}
\usepackage{dcolumn}
\usepackage{bm}
\usepackage{amsmath}  
\usepackage{algpseudocode}

\newcommand{\R}{{\mathbb R}}
\newcommand{\A}{A}
\newcommand{\E}{\mathcal{E}}
\newcommand{\rbar}{{\bar{r}}}
\newcommand{\C}{{\mathcal{C}}}

\newcommand{\oomega}{{\boldsymbol{\omega}}}
\newcommand{\vvarphi}{{\boldsymbol{\varphi}}}
\newcommand{\pphiasymp}{\hat{\boldsymbol{\varphi}}}
\newcommand{\phiasymp}{\hat \varphi}

\newcommand*{\change}[1]{ #1}

\title{Model Reduction for Kuramoto Models with Complex Topologies}
\author{Edward J. Hancock}
\address{ School of Mathematics and Statistics, The University of Sydney; Charles Perkins Centre, The University of Sydney}
\author{Georg A. Gottwald}%
\address{School of Mathematics and Statistics, The University of Sydney}
\email{georg.gottwald@sydney.edu.au}

\begin{document}
             
\begin{abstract}
Synchronisation of coupled oscillators is a ubiquitous phenomenon, occurring in topics ranging from biology and physics, to social networks and technology. A fundamental and long-time goal in the study of synchronisation has been to find low-order descriptions of complex oscillator networks and their collective dynamics. However, for the Kuramoto model - the most widely used model of coupled oscillators - this goal has remained surprisingly challenging, in particular for finite-size networks. Here, we propose a model reduction framework that effectively captures synchronisation behaviour in complex network topologies. This framework generalises a collective coordinates approach for all-to-all networks [Gottwald (2015) Chaos 25, 053111] by incorporating the graph Laplacian matrix in the collective coordinates. We first derive low dimensional evolution equations for both clustered and non-clustered oscillator networks. We then demonstrate in numerical simulations for Erd\H{o}s-R\'enyi (ER) networks that the collective coordinates capture the synchronisation behaviour in both finite-size networks as well as in the thermodynamic limit, even in the presence of interacting clusters. 
\end{abstract}



\maketitle


\section{Introduction}\label{sec:intro}
\noindent
The dynamics of interacting oscillators in complex networks is a ubiquitous model in many fields of science and engineering with examples ranging from the activity of the brain \cite{SheebaEtAl08,BhowmikShanahan12} to the functioning of power grids \cite{FilatrellaEtAl08}. A hallmark of the observed dynamics is the emergence of collective synchronised behaviour of these oscillators \cite{Kuramoto,OsipovEtAl,PikovskyEtAl,AcebronEtAl05,ArenasEtAl08,DoerflerBullo14,RodriguesEtAl16}. The ability of a network to synchronise and the nature of the transition to synchronisation depends strongly on the network topology and on the distribution of the native frequencies. The existence of a synchronised state suggests that it is possible to reduce the complexity of these potentially high-dimensional dynamical systems to just a few degrees of freedom describing the collective behaviour. Recent years have seen some progress in this direction for the widely used Kuramoto model \cite{OttAntonson08,PikovskyRosenblum08,MarvelEtAl09,MartensEtAl09,PikovskyRosenblum11,Gottwald15,Gottwald17}. Most methods, however, assume the case of a network with infinitely many oscillators. Recently, a model reduction based on collective coordinates was introduced which does not rely on the thermodynamic limit \cite{Gottwald15}. It has since been used to derive optimal synchronisation design strategies and optimal synchrony network topologies \cite{PintoSaa15,BredeKalloniatis16}. In a stochastic Kuramoto model it allowed for the quantitative description of finite-size effects, in particular the collective diffusion of the mean phase \cite{Gottwald17}. This collective coordinate approach employs a judiciously chosen ansatz function which approximates the phases of the oscillators as a function of their native frequencies. The temporal evolution of these synchronisation modes is given by the collective coordinates. However, this reduction methodology has only been formulated for the case of all-to-all coupling networks and to annealed complex networks, where averages over network configurations were performed. 

In this paper, we propose a model reduction framework that effectively captures synchronisation behaviour in complex network topologies. To do so, we introduce two main advances. First, by incorporating the network's graph Laplacian in the collective coordinate ansatz, we generalise the original approach to complex network topologies. Using the novel ansatz, we derive low dimensional evolution equations for arbitrary network topologies, both for networks with a single synchronised cluster and for networks consisting of several interacting partially synchronised clusters. Second, we present \change{a method} to identify those oscillators which do not participate in the collective behaviour - an issue relevant for intermediate coupling strengths where partial synchronisation occurs. Whereas, identifying those non-participating oscillators was straight forward in the all-to-all coupling network where they are the nodes associated with native frequencies of largest absolute value, this simple rule cannot be extended to arbitrary network topologies. The collective coordinate ansatz in conjunction with the oscillator identification method, constitutes a novel framework to effectively approximate the collective behaviour of finite complex networks with arbitrary topology. \change{To demonstrate the ability of the proposed framework to capture the collective behaviour of coupled oscillators we perform numerical simulations for Erd\H{o}s-R\'enyi networks, including a case with two interacting clusters.}

The paper is organised as follows. Section~\ref{sec:model} briefly introduces the Kuramoto model. Section~\ref{sec:coll} presents the collective coordinate framework for general network topologies. Section~\ref{sec:results} presents numerical simulations \change{for the particular case of} an Erd\H{o}s-R\'enyi networks. We conclude in Section~\ref{sec:out} with a discussion of our results and an outlook.


\section{Model}
\label{sec:model}
A widely used model for the description of interacting oscillators is the Kuramoto model \cite{Kuramoto,OsipovEtAl,PikovskyEtAl,Strogatz00,AcebronEtAl05,ArenasEtAl08,DoerflerBullo14,RodriguesEtAl16}. The Kuramoto model governs the dynamics of the phases $\varphi_i$ of $N$ interacting phase oscillators with native frequencies $\omega_i$ and is given by
\begin{align}
\dot{\varphi}_i=\omega_i+\frac{K}{N}\sum_{i=1}^N a_{ij}\sin(\varphi_j-\varphi_i).
\label{e.kuramoto}
\end{align}
Here $K$ denotes the coupling strength and $\A=[a_{ij}]$ is the adjacency matrix encoding the topology of the network. We assume here that the network is not directed with a symmetric unweighted adjacency matrix $A$ with $a_{ij}=a_{ji}=1$ if there is an edge between oscillators $i$ and $j$, and $a_{ij}=0$ otherwise. The degree of a node $d_i$ is then given by $d_i=\sum_j a_{ij}$. We introduce for later the graph Laplacian
\begin{align}
L = D-A,
\label{e.GL}
\end{align}
with degree matrix  $D={\rm{diag}}(d_1,d_2,\cdots,d_N)$. Note that the graph Laplacian of a fully connected network has a single zero-eigenvalue with eigenvector ${\bm{1}}_{N}$. We assume that the native frequencies are distributed according to some distribution $g(\omega)$ and have, without loss of generality, mean zero, i.e. ${\bm{1}}_{N}^T{\oomega}=0$ where $\oomega=(\omega_1,\cdots,\omega_N)^T$ denotes the vector of natural frequencies.\\ 

Typically, once the coupling strength is sufficiently strong with $K>K_c$ for some critical coupling strength $K_c$, synchronisation occurs in the sense that the oscillators become locked to their mutual mean frequency $\bar\Omega=\tfrac{1}{N}\sum_{i=1}^N \omega_i$ and their phases become localised about their mean phase \cite{Kuramoto,OsipovEtAl,Strogatz00}. This type of synchronous behaviour is known as global synchronisation and is characterised by a globally attracting manifold on which the dynamics settles \cite{Crawford94}. The level of synchronisation is often characterised by the order parameter \cite{Kuramoto}
\begin{align}
\label{e.r}
r(t)=\frac{1}{N}\big\lvert \sum_{j=1}^Ne^{i\varphi_j(t)}\big\rvert
\end{align} 
with $0\le r \le 1$. In practice, the asymptotic limit of the order parameter 
\begin{align}
\label{e.rbar}
{\bar{r}} = \lim_{T\to\infty}\frac{1}{T}\int_{T_0}^{T_0+T} r(t)\, dt 
\end{align}
is estimated whereby $T_0$ is chosen sufficiently large to eliminate eventual transient dynamics. 

In the case of full synchronisation with $\varphi_i(t)=\varphi_j(t)$ for all pairs $i,j$ and for all times $t$ we obtain $\rbar=r=1$. In the case where all oscillators behave independently with random initial conditions, $\rbar = \mathcal{O}(1/\sqrt{N})$ indicates incoherent phase dynamics; values in between indicate partial coherence. 


\section{Collective Coordinates}
\label{sec:coll}

In this section, we generalise the collective coordinate methodology introduced in \cite{Gottwald15}. We first present the collective coordinate framework for the situation when there is a single cluster of oscillators which tends to mutual synchronisation; we then set out to present the collective coordinate framework which takes into account the situation when several individually but not mutually synchronised clusters interact. In the collective coordinate framework the phases of the $N$ oscillators are expressed via an ansatz function 
\begin{align}
\varphi_i(t) = {\Phi}_i(\alpha_1(t),\cdots,\alpha_n(t);\omega,\A)
\label{e.cc00}
\end{align}
for $i=1,\cdots,N$ and $n\ll N$ collective coordinates ${{\alpha}}_j$. The temporal evolution of the $N$ phase variables $\varphi_i$ is then described by $n$ collective coordinates $\alpha_j$. This reduces an $N$ dimensional system to an $n$ dimensional one. For all-to-all networks with $a_{ij}=1$ for all $i,j$ the ansatz $\varphi_i(t)=\Phi_i(t)$ with
\begin{align}
\Phi_i(t) = \alpha(t) \, \omega_i
\label{e.cc_alltoall}
\end{align}
was proposed in \cite{Gottwald15}. In the case of a bimodal frequency distribution, which allows for interacting partially synchronised clusters, one has to introduce an additional collective coordinate to capture this interaction. The ansatz (\ref{e.cc_alltoall}) was numerically verified and can be motivated in the limit of large coupling strength $K\gg 1$. The stationary Kuramoto model (\ref{e.kuramoto}) can be cast as $\omega_i = - K r \sin(\psi-\varphi_i)$ introducing the mean phase $\psi$ \cite{Kuramoto}. Expanding $\varphi_i=\psi +\arcsin(\omega_i/(rK))$ in $1/K$ for large coupling strength yields up to first order $\varphi_i = \psi + \omega_i/(rK)$. Since the Kuramoto model is invariant under constant phase shifts we may set $\psi=0$ leading to the collective coordinate ansatz (\ref{e.cc_alltoall}) \footnote{For stochastic Kuramoto models, the mean phase experiences non-trivial diffusive behaviour which can also be captured by the collective coordinate framework, see \cite{Gottwald17}}. The evolution equations for the collective coordinates are determined by minimising the error accrued by restricting the solutions to be of the form (\ref{e.cc_alltoall}); the reader is referred to \cite{Gottwald15} for details.


\subsection{A single synchronising cluster with complex topology}
\label{sec.single}
To devise an appropriate ansatz for general network topologies, we again focus on the strongly synchronised state for large $K$. In the asymptotic limit $K\to \infty$ the globally synchronised state  $\varphi_i=\varphi_j={\rm{const}}$ can be approximated (ignoring a constant mean phase $\psi$) via linearisation as
\begin{align}
\pphiasymp = \frac{N}{K}L^+\oomega,
\label{e.phiasy}
\end{align}
where $L^+$ denotes the pseudo-inverse of the graph Laplacian (\ref{e.GL}) (see, for example, \cite{Golub}). \change{Note that the kernel mode ${\bm{1}}_{N}$ of the graph Laplacian is associated with the invariance of the Kuramoto model with addition of a mean phase to $\vvarphi$.} \change{Assuming that the mode $\pphiasymp $ dominates the dynamics} this suggests the following ansatz with collective coordinate $\alpha(t)$:
\begin{align}
\Phi =  \alpha(t)\, \pphiasymp. 
\label{e.cc_gen}
\end{align}
Note that $L^+{\bm{1}}_{N} = 0$ and ${\bm{1}}_{N}^TL^+ \oomega=0$ for any native frequency vector $\oomega$. We remark that for all-to-all networks we have $L=N I_n-{\bm{1}}_{N} {\bm{1}}_{N}^T$ and the ansatz (\ref{e.cc_gen}) reduces to the ansatz (\ref{e.cc_alltoall}) with $\alpha$ being scaled with $1/K$ \footnote{Note that $L$ has a single zero eigenvalue with corresponding eigenvector $V_1$ satisfying $V_1^T\omega=0$, and $N-1$ repeated eigenvalues $\lambda=N$. Using an eigenvalue decomposition, write $L^+\omega=V D^+V^T\omega=\lambda^{-1} \omega$ which implies (\ref{e.cc_alltoall})}. 

We now follow \cite{Gottwald15} to determine the temporal evolution equations for the collective coordinate $\alpha$. Inserting the ansatz (\ref{e.cc_gen}) into the Kuramoto model we obtain the error made by the collective coordinate ansatz (\ref{e.cc_gen})
\begin{align*}
\E_i=\dot{\alpha}{\phiasymp}_i-\omega_i-\frac{K}{N}\sum_{\substack{j=1\\j\ne i}} a_{ij}\sin(\alpha(\phiasymp_j-\phiasymp_i)).
\end{align*}
We wish to \change{minimise the error ${\boldsymbol{\E}}$ made by restricting the solution space to the one-dimensional subspace spanned by $\pphiasymp$, or equivalently to maximise the degree to which our collective coordinates are capable of capturing the dynamics of the full Kuramoto model.} \change{This is achieved by assuring that the error is orthogonal to $\pphiasymp$.} Setting $\sum_i\E_i\phiasymp_i = 0$, we obtain an evolution equation for the collective coordinates
\begin{align}
\dot{\alpha}=\frac{K}{N}\frac{\pphiasymp^{T} L\,\pphiasymp}{\pphiasymp^{T} \pphiasymp}+\frac{1}{\pphiasymp^{T} \pphiasymp}\frac{K}{N}\sum_{i,j} \phiasymp_i a_{ij}\sin(\alpha(\phiasymp_j-\phiasymp_i)).
\label{e.alpha_raw}
\end{align}
Upon rescaling time such that $t=T_s\tau$ with 
\begin{align*}
T_s =\frac{N}{K} \frac{\pphiasymp^{T}\pphiasymp}{\pphiasymp^{T} L\, \pphiasymp},
\end{align*}
the evolution equation (\ref{e.alpha_raw}) for the collective coordinate $\alpha$ simplifies to
\begin{equation}
\dot{\alpha}=1+\frac{1}{\pphiasymp^{T}L\pphiasymp}\sum_{i,j} \phiasymp_i a_{ij}\sin(\alpha(\phiasymp_j-\phiasymp_i)).
\label{e.alpha}
\end{equation}
Equilibrium solutions $\alpha^\star$ with $\dot \alpha^\star=0$ correspond to the synchronised state and the transition to synchronisation appears at $K=K_c$ which is the smallest $K$ such that (\ref{e.alpha}) supports stable equilibrium solutions.\\ 

\begin{figure}
\begin{center}
\includegraphics[width=0.5\textwidth, height=0.25\textheight]{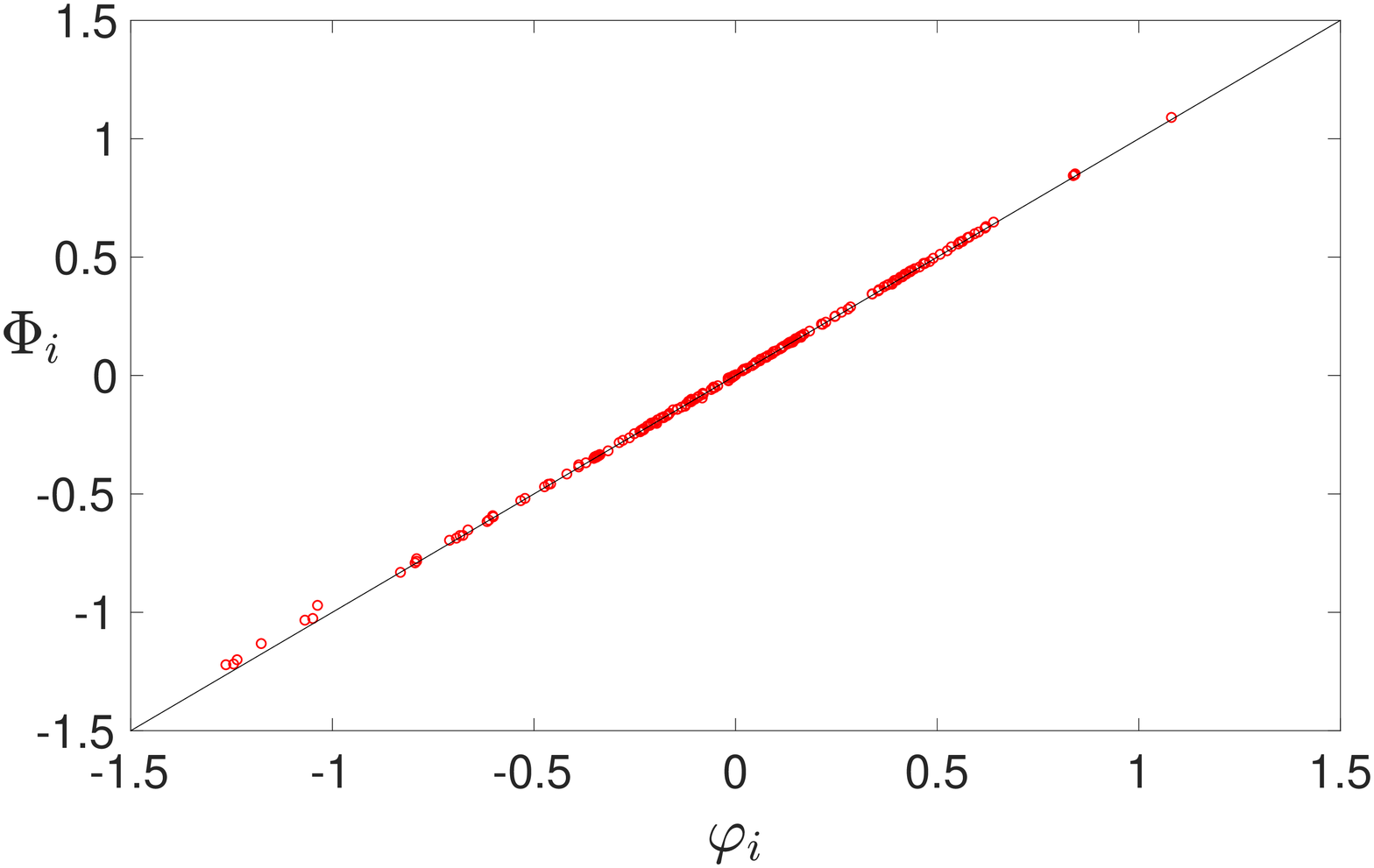}
\end{center}
\caption{Snapshot of the phases $\varphi_i$ obtained from simulating the full Kuramoto model (\ref{e.kuramoto}) against the collective coordinate ansatz $\Phi_i$ (\ref{e.cc_gen}) for $K=180$, with $\alpha=1.03$ obtained as the stationary solution of (\ref{e.alpha}), for a small-world topology with $N=200$ oscillators, each with $2$ neighbours and a rewiring probability $p=0.3$, and native frequencies drawn from a normal distribution $\mathcal{N}(0,0.1)$. The corresponding value of the order parameter is $\rbar=0.92$. The continuous line indicates perfect correspondence between the ansatz and the observed phases.}
\label{f.complexnetwork_ansatz}
\end{figure}

Figure~\ref{f.complexnetwork_ansatz} provides a numerical illustration of the validity of the collective coordinate ansatz (\ref{e.cc_gen}) where we plot the actually observed phases against the collective coordinate ansatz (\ref{e.cc_gen}) for a small-world network \cite{WattsStrogatz98} and native frequencies drawn from a normal distribution. The oscillators are clearly well described by the collective coordinate ansatz. The agreement of the phases with the collective coordinate ansatz becomes better for increasing coupling strength. We remark that upon decreasing the coupling strength, not all oscillators are able to synchronise and only a subset of the $N$ oscillators will satisfy the ansatz (\ref{e.cc_gen}). \change{This is addressed in the following.}\\

Particular choices of frequency distributions or the presence of topological communities within the network may not allow for the global synchronisation of all $N$ oscillators at a given coupling strength $K$. Instead one observes one or several partially or locally synchronised clusters, with possible complex interactions between them. One example, already discussed in \cite{Gottwald15}, is the Kuramoto model with a unimodal frequency distribution, where the transition to synchronisation is a second-order phase transition~\cite{Kuramoto,OsipovEtAl}, and not all oscillators participate in the collective synchronised state: As the coupling strength is increased from zero, at some critical strength $K=K_l$ a few oscillators perform collective behaviour and mutually synchronise. Increasing the coupling strength then allows increasingly more oscillators to become entrained to the synchronised state until global synchronisation sets in at $K=K_c$. Hence for coupling strength $K_l\le K<K_c$ which allows for local synchronisation we cannot expect to find an equilibrium solution of (\ref{e.alpha}). To capture this local synchronisation for a given coupling strength $K$ within the collective coordinate framework, we assume that those oscillators that can mutually synchronise will do so. This suggests the ansatz (\ref{e.cc_gen}) $\varphi_i=\alpha(t)\,\phiasymp_i$ for $i\in \mathcal{C}_m$ where we denote by $\mathcal{C}_m$ the largest set of nodes which can synchronise. The collective coordinate ansatz is then applied only to nodes in $\mathcal{C}_m$ and the condition on the minimisation of the error reads $\sum_{i\in\mathcal{C}_m}\E_i \phiasymp_i^{(m)}=0$, leading to 
\begin{align}
\dot{\alpha}=1+\frac{1}{{\pphiasymp}^{(m)^T} L_{m}\, \pphiasymp^{(m)}}\sum_{i\in \C_m}\sum_{j\in \C_m}\phiasymp_i^{(m)} a_{ij}\sin(\alpha(\phiasymp_j^{(m)}-\phiasymp_i^{(m)}))
\label{e.alpha_c}
\end{align}
with the asymptotic mode
\begin{align}
\pphiasymp^{(m)}=\frac{N}{K}L_m^+\oomega^{(m)},
\label{e.phiasy_m}
\end{align}
where $\oomega^{(m)}$ denotes the native frequencies of nodes in the set $\mathcal{C}_m$ and $L_m$ is the graph Laplacian of the network consisting only of nodes in the set $\mathcal{C}_m$ (cf. (\ref{e.cc_gen})). The set $\mathcal{C}_m$ is determined such that its cardinality $N_m$ is the the largest possible size such that (\ref{e.alpha_c}) admits a stable equilibrium solution $\dot \alpha^\star=0$. In the case of an all-to-all coupling network, this is readily achieved by excluding successively those nodes with native frequencies with the largest absolute frequencies (see \cite{Gottwald15}). In the case of arbitrary connected complex networks we are not aware of any computationally efficient way to test for the largest set of nodes allowing for stable stationary equilibrium solutions $\alpha^\star$. \change{We propose here a dynamical criterion to identify those non-entrained nodes} by linearising the Kuramoto model (\ref{e.kuramoto}) around an equilibrium solution $\alpha^\star$, and consider the linearised matrix
\begin{align}
\left(L_{\rm{lin}}\right)_{ij} = 
\begin{cases}
a_{ij}\cos(\alpha^\star(\phiasymp_j^{(m)}-\phiasymp_i^{(m)})),\quad i\ne j\\
-\sum_k a_{ik}\cos(\alpha^\star(\phiasymp_k^{(m)}-\phiasymp_i^{(m)})),\quad i= j\\
\end{cases}.
\label{e.Llin}
\end{align}
For a stable system, $L_{\rm{lin}}$ has one zero eigenvalue and $N-1$ negative eigenvalues. The dynamics becomes linearly unstable when an eigenvalue of $L_{\rm{lin}}$ becomes positive. \change{We remark that this is not addressing the stability of the collective coordinate system (\ref{e.alpha}), but rather the stability of the approximation $\vvarphi = \alpha(t)\pphiasymp$ within the full Kuramoto model.}\\  
\change{
A set of nodes $\mathcal{C}_m$ needs to satisfy two conditions: first equilibrium solutions $\alpha^\star$ of (\ref{e.alpha_c}) need to exist and second they need to be linearly stable. For sufficiently large values of the coupling strength $K$ linearly stable equilibrium solutions can be found corresponding to global synchronisation with $N_m=N$ (note that in this case (\ref{e.alpha_c}) and (\ref{e.alpha}) are identical). Decreasing the coupling strength $K$ in small increments $\delta K$ while keeping the number of oscillators fixed for each step, we reach a coupling strength $K^\prime$ where either equilibrium solutions of (\ref{e.alpha_c}) cease to exist or where the equilibrium solution turns linearly unstable. In the latter case we consider the eigenvector $\hat v$ of $L_{\rm{lin}}$ corresponding to the positive eigenvalue. The set of nodes to be excluded from the collective coordinate ansatz is determined by the components of $\hat v$ with large absolute values. Algorithmically, this is achieved by ordering the components of the eigenvector $\hat v$ and determining the largest difference or gap between neighbouring components. The network is then partitioned between those nodes above and below the largest gap, where the group with less elements is discarded. 
We remark that if this procedure excludes nodes such that the remaining network is disconnected, we choose the largest connected network within this set of nodes.\\}
\change{
If upon decreasing the coupling strength $K$ in small increments $\delta K$ while keeping the number of oscillators fixed for each step, we reach a coupling strength $K^\prime$ such that no equilibrium solution exists, then $L_{\rm{lin}}$ is evaluated around the last equilibrium solution $\alpha^\star$ at $K=K^\prime+\delta K$, and the nodes to be excluded from the synchronised set are again determined from the eigenvector $\hat v$ associated with the largest non-zero eigenvalue.}
\change{This is justified when the eigenvalues depend continuously on $K$. A computationally more costly procedure would be to use bisection to find a value of $K$ for which an equilibrium solution exists and then determine its unstable eigenvector.}

In the following we present the collective coordinate framework when there are more than one locally synchronised cluster.
 

\subsection{Interacting locally synchronised clusters with complex topology}
\begin{figure}
\begin{center}
\includegraphics[width=0.5\textwidth, height=0.25\textheight]{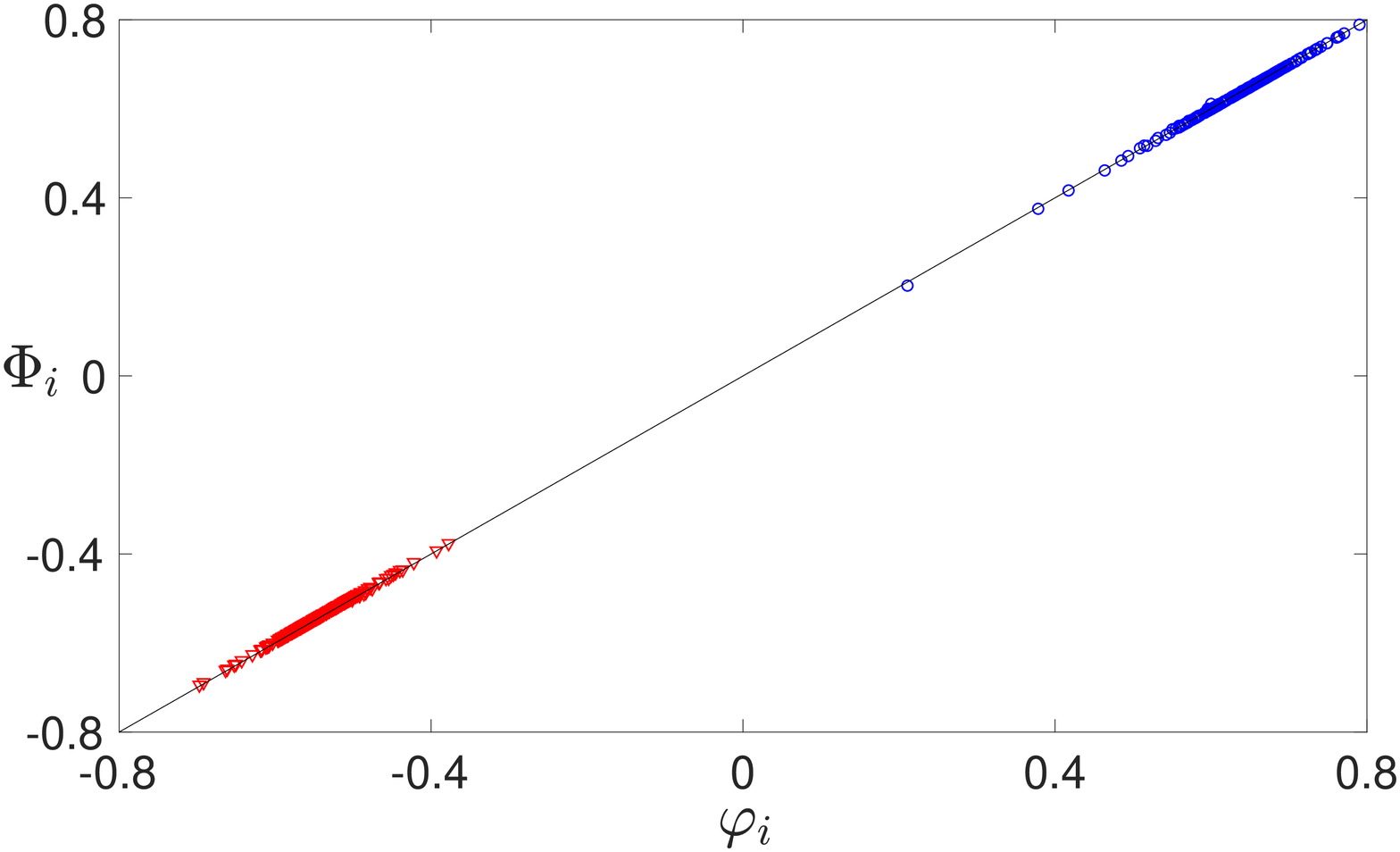}
\end{center}
\caption{Snapshot of the phases $\varphi_i$ obtained from simulating the full Kuramoto model (\ref{e.kuramoto}) against the collective coordinate ansatz $\Phi_i^{(1,2)}$ (\ref{e.cc_cluster}), with $\alpha_1=1.003,\alpha_2=1.006, f_1=-0.095$ and $f_2=0.11$ obtained from (\ref{eq:red_model_m_alpha}) and (\ref{eq:red_model_m_f}). The network is an ER network with $N=500$ which consists of two topological clusters with only a few inter-cluster connections. The two clusters are labelled red and blue, respectively. 
The native frequencies are drawn from a normal distribution $\mathcal{N}(0,0.02)$. The snapshot is taken at $K=160$ and the corresponding value of the order parameter is $\rbar=0.83$. The continuous line indicates perfect correspondence between the ansatz and the observed phases.}
\label{f.complexnetwork_ansatz_cluster}
\end{figure}

We now set out to formulate the collective coordinate ansatz allowing for the interaction between several locally synchronised clusters. Let us consider that there are one or several sets of nodes ${\mathcal{C}}_m$ with $m=1,\cdots,M$, each of size $N_{m}$ which exhibit localised collective behaviour within their respective sets. We write the Kuramoto model (\ref{e.kuramoto}) for the phases of nodes in the $m$th cluster, $\varphi^{(m)}\in\R^{N_m}$, with native frequencies $\omega_i^{(m)}$ as
\begin{align}
\dot{\varphi}_i^{(m)}=\omega_i^{(m)}+\frac{K}{N}\sum_{k=1}^{M}\sum_{j\in \C_k} a_{ij}\sin(\varphi^{(k)}_j-\varphi^{(m)}_i),
\label{e.kuramoto_cluster}
\end{align}
for $i\in \C_m$. To capture the collective behaviour within each set, the collective coordinate approach is then restricted to each set individually. We introduce collective coordinates $\alpha_m(t)$ to describe the collective behaviour within a cluster $\mathcal{C}_m$ and collective mean phase coordinates $f_m(t)$ to account for the inter-cluster dynamics \footnote{See \cite{Gottwald15} for an example of two interacting clusters in an all-to-all coupling network with a bimodal frequency distribution.}. \change{As discussed in Section~\ref{sec.single}, a single cluster in isolation is described by $\alpha_m(t)\, \pphiasymp^{(m)}$ with $\pphiasymp^{(m)}=(N/K)L_m^+\oomega^{(m)}$ where $L_m=D_m-A_m$ denotes the graph Laplacian for the $m$th cluster with the cluster's adjacency matrix $A_m=[a_{ij}]$ restricted to $i,j\in\mathcal{C}_m$ and associated degree matrix $D_m$ (cf. (\ref{e.phiasy_m})). Since $\pphiasymp^{(m)}$ is agnostic about the nodes from other clusters $\mathcal{C}_k$ with $k\neq m$, the asymptotic mode $\pphiasymp^{(m)}$ does not correspond to the asymptotic state $\pphiasymp$ when projected onto the set of nodes in the set $\mathcal{C}_m$. We denote the difference by
\begin{align}
\Delta\pphiasymp^{(m)}
=
\pi_m(\pphiasymp)-\pphiasymp^{(m)},
\end{align}
where $\pi_m(\pphiasymp)$ denotes the projection of the global asymptotic solution $\pphiasymp$ onto the set of nodes in the $m$th cluster $\mathcal{C}_m$. 
Recall that $ f_m(t)\,{\bm{1}}_{N_m}$ represents a mean phase added to the $m$th cluster which would not change the solution if the cluster were in isolation. Mathematically this is reflected by ${\bm{1}}_{N_m}$ being in the null space of $L_m$. We now represent the collective effect of the other clusters onto the $m$th cluster by a change in its overall phase. We therefore restrict the solution to lie within the affine subspace spanned by the dominant modes $\pphiasymp^{(m)}$ and ${\bm{1}}_{N_m}$ and write
\begin{align}
\Phi^{(m)}=\Delta\pphiasymp^{(m)} + \alpha_{m}(t)\,\pphiasymp^{(m)} + f_{m}(t)\, {\bm{1}}_{N_m} ,
\label{e.cc_cluster}
\end{align}
which for the fully synchronised case $K\to\infty$ recovers the asymptotic state $\pphiasymp$ for one single globally synchronised cluster with $\alpha_m\to 1$ and $f_m=f_l$ for all $m,l$ (with abuse of notation with respect to the general ansatz (\ref{e.cc00})). 
}
The collective coordinates $\alpha_m(t)$ describe the intra-cluster dynamics whereas the collective coordinate $f_m(t)$ describe the inter-cluster dynamics. Note that the inter-cluster collective coordinates satisfy $\sum_{m=1}^M N_m f_m = 0$.\\ Figure~\ref{f.complexnetwork_ansatz_cluster} shows a snapshot of the phases for an ER network consisting of two \change{well-specified topological clusters with strong intra-cluster and weak inter-cluster connectivity (details are given below in Section~\ref{sec:results})}. We show the actual phases obtained from a numerical simulation of the full Kuramoto model (\ref{e.kuramoto}) and the results of the collective coordinate ansatz (\ref{e.cc_cluster}) for the two clusters. \change{The phases are remarkably well reproduced. We remark that the colours of the nodes labelling the two clusters in Figure~\ref{f.complexnetwork_ansatz_cluster} represent the two sets of nodes identified by the most unstable eigenvector $\hat v$ of $L_{\rm{lin}}$, which in this case correctly identifies the two topological clusters.}\\

Inserting the collective coordinate ansatz (\ref{e.cc_cluster}) into the Kuramoto model (\ref{e.kuramoto_cluster}), we obtain the error for $i\in \mathcal{C}_m$
\begin{align}
\E_i^{(m)}=\dot{\alpha}_m \phiasymp^{(m)}_i +\dot{f}_m-\omega_i^{(m)} 
-\frac{K}{N}\sum_{\substack{k=1}}^{M}\sum_{j\in \C_k} a_{ij}\sin(\Phi^{(k)}_j-\Phi^{(m)}_i).
\end{align}
\change{Again we seek to minimise the error made by restricting the solution space to the two-dimensional subspace spanned by $\pphiasymp^{(m)}$ and ${\bm{1}}_{N_m}$, i.e. we require the error to be orthogonal to $\pphiasymp^{(m)}$ and ${\bm{1}}_{N_m}$.} Setting $\sum_{i\in \C_m} \E_i^{(m)} \phiasymp_i^{(m)}=0$ and $\sum_{i\in \C_m}  \E_i^{(m)}=0$ for all $m=1,\cdots,M$, this yields the evolution equations for the intra-cluster collective variable $\alpha_m$ with
\begin{align}
\dot{\alpha}_m
=
\frac{K}{N}\frac{\pphiasymp^{(m)^T} L_m \pphiasymp^{(m)} }{\pphiasymp^{{(m)}^T} \pphiasymp^{(m)} }
+
\frac{1}{\pphiasymp^{{(m)}^T} \pphiasymp^{(m)}}\frac{K}{N}\sum_{\substack{k=1}}^{M}\sum_{j\in \C_k}\sum_{i\in \C_m} \phiasymp_i^{(m)} a_{ij}\sin(\Phi^{(k)}_j-\Phi^{(m)}_i)
\label{eq:red_model_m_alpha}
\end{align}
and for the inter-cluster variable $f_m$ with
\begin{align}
\dot{f}_m=\Omega_c^{(m)}+\frac{1}{N_m}\frac{K}{N}\sum_{\substack{k=1\\k\ne m}}^{M}\sum_{j\in \C_k}\sum_{i\in \C_m} a_{ij}\sin(\Phi^{(k)}_j-\Phi^{(m)}_i) ,
\label{eq:red_model_m_f}
\end{align}
where $\Omega_c^{(m)} = \sum_{i\in\C_m}\omega_i^{(m)}/N_m$ is the mean of the native frequencies in the $m$th cluster. We remark that for a single cluster $M=1$ the evolution equations (\ref{eq:red_model_m_alpha})--(\ref{eq:red_model_m_f}) reduce to (\ref{e.alpha_c}) with ${\dot{f}}_1 = 0$, in accordance with the invariance of the Kuramoto model with respect to adding a constant mean phase.


\section{Numerical results}
\label{sec:results}

We test the proposed methodology on unweighted Erd\H{o}s-R\'enyi networks. In an Erd\H{o}s-R\'enyi network nodes are connected independently with probability $p$ and with Poisson-distributed degrees $d_j$ with mean degree $d=pN$. We choose here $p=0.05$ throughout. We present results for randomly distributed native frequencies, drawn from a distribution $g(\omega)$. In particular, we consider uniformly distributed native frequencies on the interval $[-1,1]$ with distribution
\begin{align}
g(\omega)=0.5 ,
\end{align}
and normally distributed native frequencies with 
\begin{align}
g(\omega)=\frac{1}{\sqrt{2\pi\sigma_\omega^2}} \exp\left(-\frac{\omega^2}{2 \sigma_\omega^2}\right)
\end{align}
with $\sigma_\omega^2=0.1$.\\

\begin{figure}[ht]
	\centering
	\includegraphics[width=0.5\textwidth]{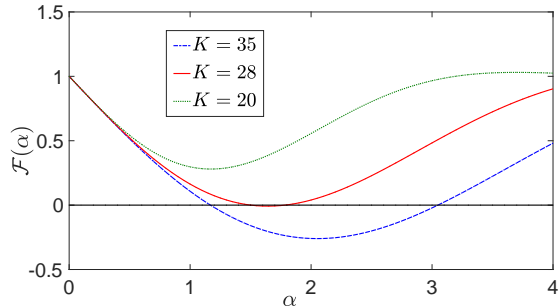}	
	\caption{Right-hand-side $\mathcal{F}(\alpha)$ of the evolution equation (\ref{e.alpha_c}) for several values of $K$ for the $N=500$ ER network with uniformly distributed native frequencies with a subcritical coupling strength $K=20<K_c$, critical coupling strength $K=28\approx K_c$ and supercritical coupling strength $K=35>K_c$.
	}
	\label{fig:ERG_uniform_SN}
\end{figure}

\change{We first study an ER network with uniformly distributed native frequencies. The collective coordinate approach identifies the nature of the bifurcation initiating the onset of synchronisation as a saddle-node bifurcation.} This is illustrated in Figure~\ref{fig:ERG_uniform_SN} where we plot the right-hand-side of the evolution equation (\ref{e.alpha_c})
\begin{align}
{\mathcal{F}}(\alpha) =
1+\frac{1}{{\pphiasymp}^{(l)^T} L_{l}\,\pphiasymp^{(l)}}\sum_{i\in \C}\sum_{j\in \C}\phiasymp_i^{(l)} a_{ij}\sin(\alpha(\phiasymp_j^{(l)}-\phiasymp_i^{(l)}))
\label{e.Falpha}
\end{align}
as a function of $\alpha$ for coupling strength $K$ below and above the critical coupling strength $K_c=27$ as well as close to $K=K_c$. Equilibrium solutions are given by ${\mathcal{F}}(\alpha^\star)=0$. It is seen that there are no solutions for $K<K_c$ and at $K=K_c$ a pair of equilibrium solutions emerges, one being stable (the smaller one, closer to $\alpha=1$) and one being unstable.

\begin{figure}[ht]
	\centering
	\includegraphics[width=0.5\textwidth]{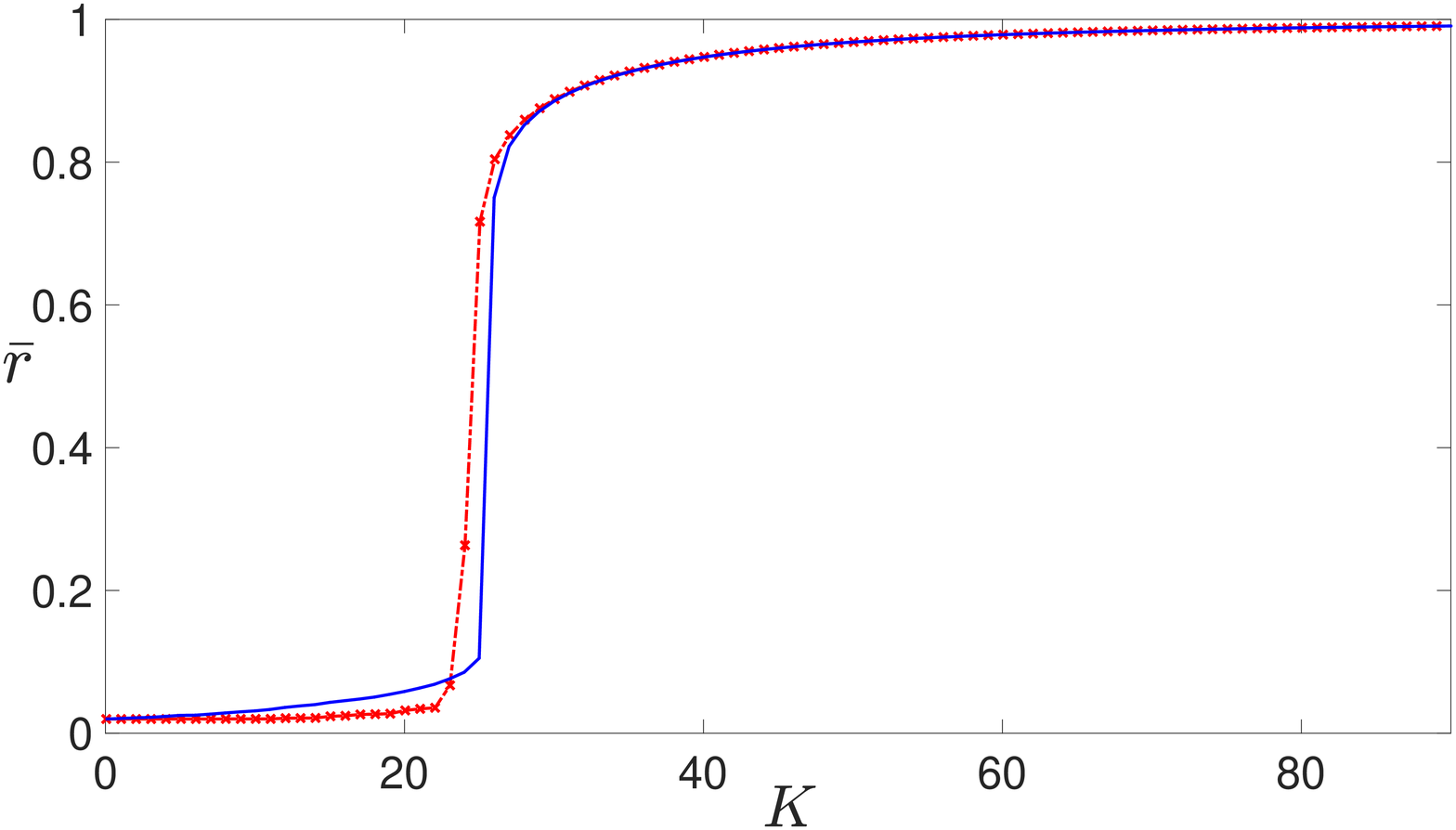}\\
	\includegraphics[width=0.5\textwidth]{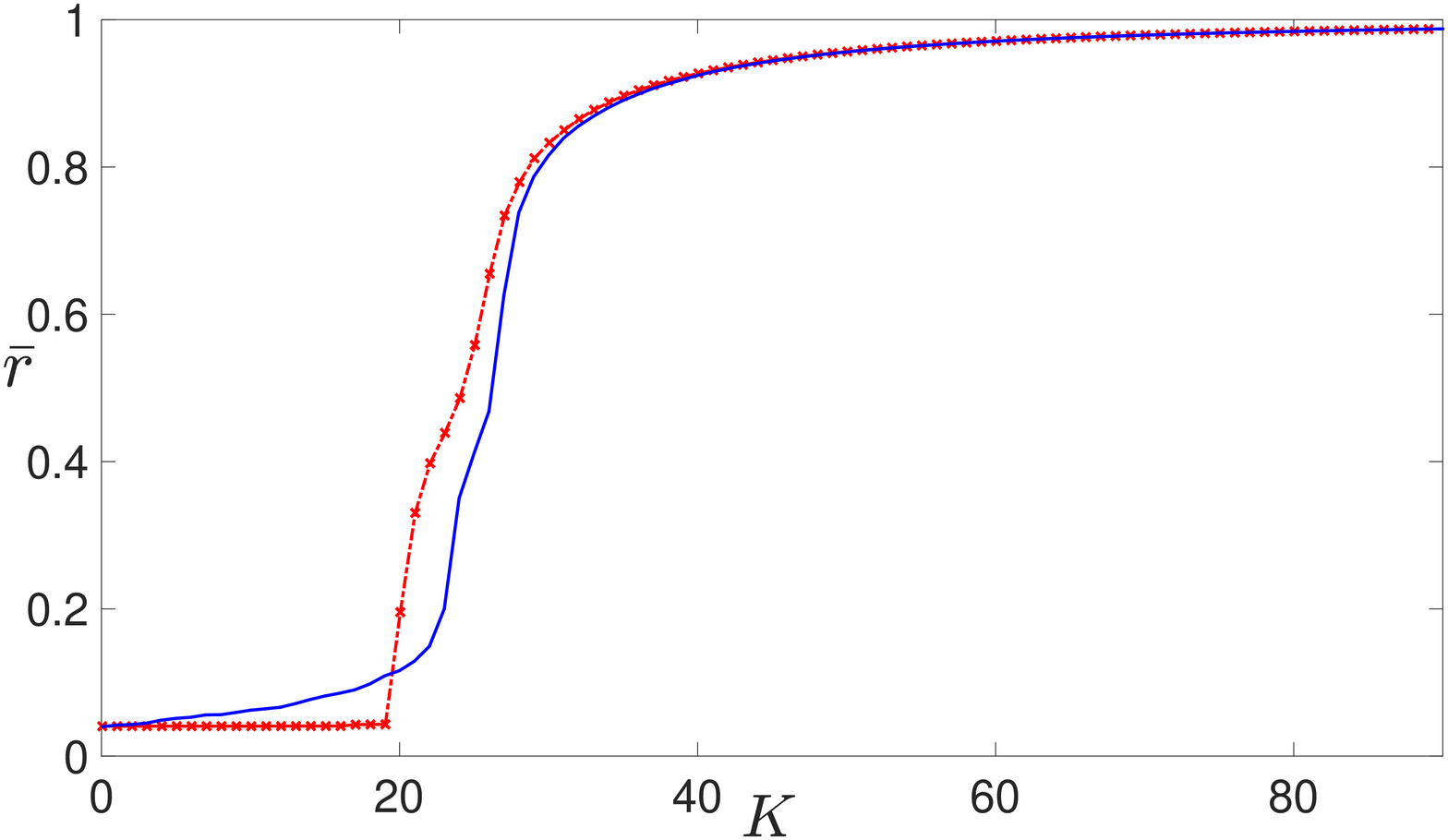}	
	\caption{Order parameter $\rbar$ as a function of the coupling strength $K$ for an ER network with uniformly distributed native frequencies. Depicted are results from a direct numerical integration of the Kuramoto model (\ref{e.kuramoto}) (continuous line, online blue) and from the collective coordinate approach (\ref{e.cc_gen}) using (\ref{e.rcc}) (crosses, online red). Top: ER network with $N=2000$ nodes. Bottom: ER network with $N=500$ nodes.}
	\label{fig:ERG_uniform}
\end{figure}

\begin{figure}[ht]
	\centering
	\includegraphics[width=0.5\textwidth]{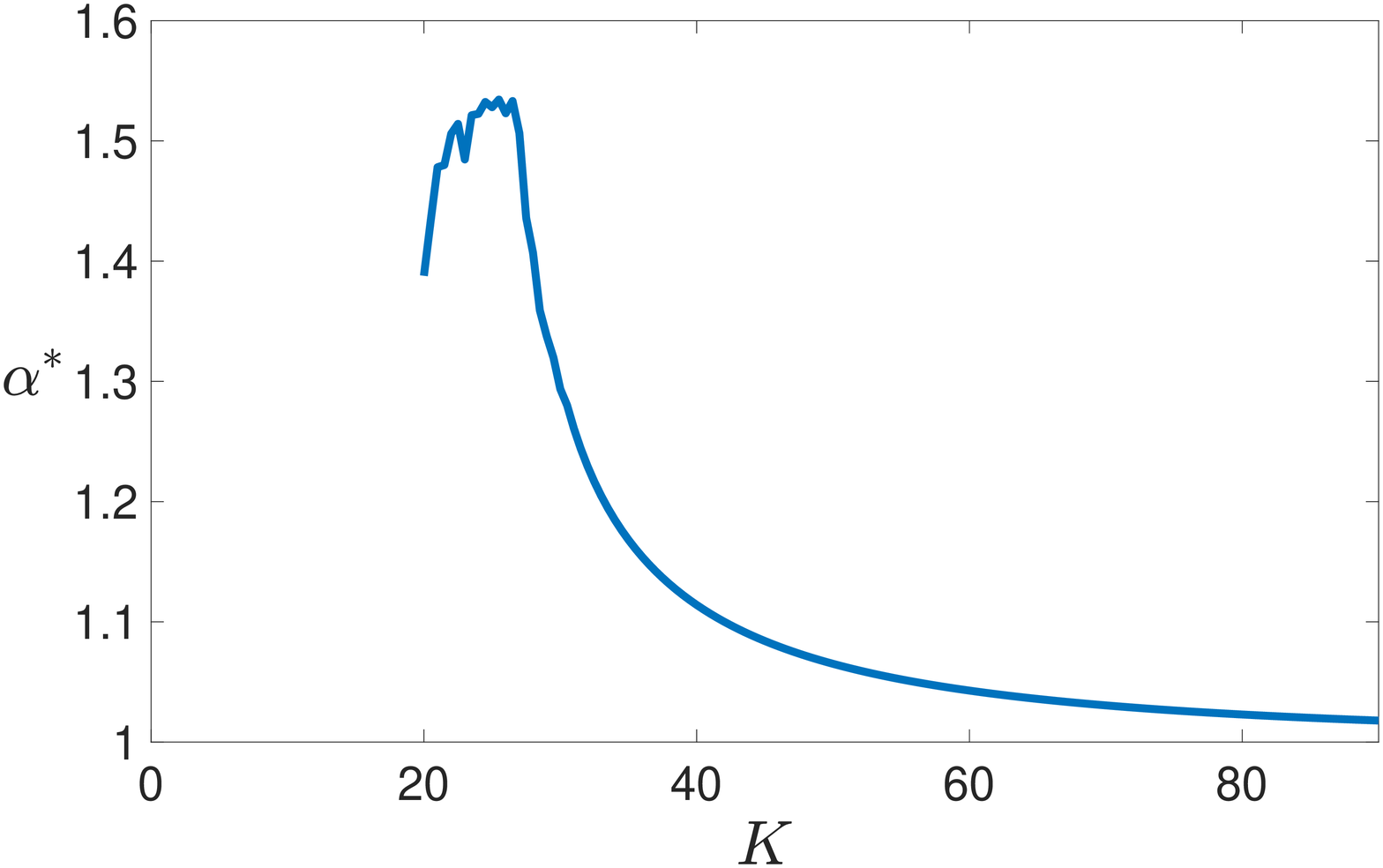}	
	\caption{Equilibrium solution $\alpha^\star$ for the $N=500$ ER network with uniformly distributed native frequencies for the same network as in Figure~\ref{fig:ERG_uniform}. The solution was obtained solving  (\ref{e.alpha_c}) for a set of nodes $\mathcal{C}_l$ consisting of linearly stable nodes according to (\ref{e.Llin}). We only plot the equilibrium solution which is closest to $\alpha=1$ from the two stationary solutions of (\ref{e.alpha_c}).  
	}
	\label{fig:ERG_uniform_alpha}
\end{figure}

We show in Figure~\ref{fig:ERG_uniform} the order parameter $\bar{r}$ as a function of the coupling strength $K$ for two networks with sizes $N=2000$ and $N=500$, respectively. The figure shows a comparison of the order parameter as calculated from a long simulation of the full Kuramoto model (\ref{e.kuramoto}) and as estimated by the collective coordinate ansatz (\ref{e.cc_gen}) where $\alpha$ is determined as the stationary solution of (\ref{e.alpha}). To solve the collective coordinate evolution equation (\ref{e.alpha}) for stationary solutions $\alpha=\alpha^\star$, we discarded any nodes corresponding to unstable eigenvectors of the linearisation matrix $L_{\rm{lin}}$ as described in Section~\ref{sec.single}. Let us denote by $\mathcal{C}_l$ the set of nodes for which an equilibrium solution $\alpha^\star$ can be found which is linearly stable according to the linearised matrix $L_{\rm{lin}}$ (see (\ref{e.Llin})). \change{Figure~\ref{fig:ERG_uniform_alpha} shows the equilibrium solution $\alpha^\star$ of  (\ref{e.alpha_c}) found for a network of $N=500$ as a function of $K$.} We then calculate the order parameter $\rbar$ of the collective coordinate using
\begin{align}
\label{e.rcc}
r_{\rm{cc}}(t)=\frac{1}{N}\big\lvert\sum_{j\in\mathcal{C}_l}e^{i\alpha^\star \phiasymp_j} + \sum_{j\notin\mathcal{C}_l}e^{i\omega_j t}\big\rvert.
\end{align} 
It is seen in Figure~\ref{fig:ERG_uniform} that the collective coordinate approach works very well for the larger network with $N=2000$ and resolves the explosive transition to synchronisation near $K_c=26$, \change{corresponding to the saddle-node bifurcation}. For the smaller network with $N=500$ nodes, the collective coordinate approach captures the collective synchronisation behaviour very well for large coupling strength $K$. For smaller coupling strengths with $K<27$ the match of the order parameters is reasonable; \change{it is seen that the qualitative behaviour and the functional form of the curve $\bar r(K)$, including the concave functional behaviour near $K=24$, is well captured by the collective coordinate approximation, but the two curves are shifted by $\Delta K \approx 2$.} This delayed synchronisation of the actual Kuramoto model (\ref{e.kuramoto}), we conjecture, is due to our method not correctly identifying nodes which do not partake in the collective synchronised behaviour captured by the ansatz at a particular value of $K$. 
Furthermore, we remark that for values of the coupling strength near the onset of synchronisation the interaction between the set of partially synchronised oscillators and the non-entrained oscillators, which may themselves form small partially synchronised clusters, is not captured by the single-cluster ansatz (\ref{e.cc_gen}).

\begin{figure}[ht]
	\centering
	\includegraphics[width=0.5\textwidth]{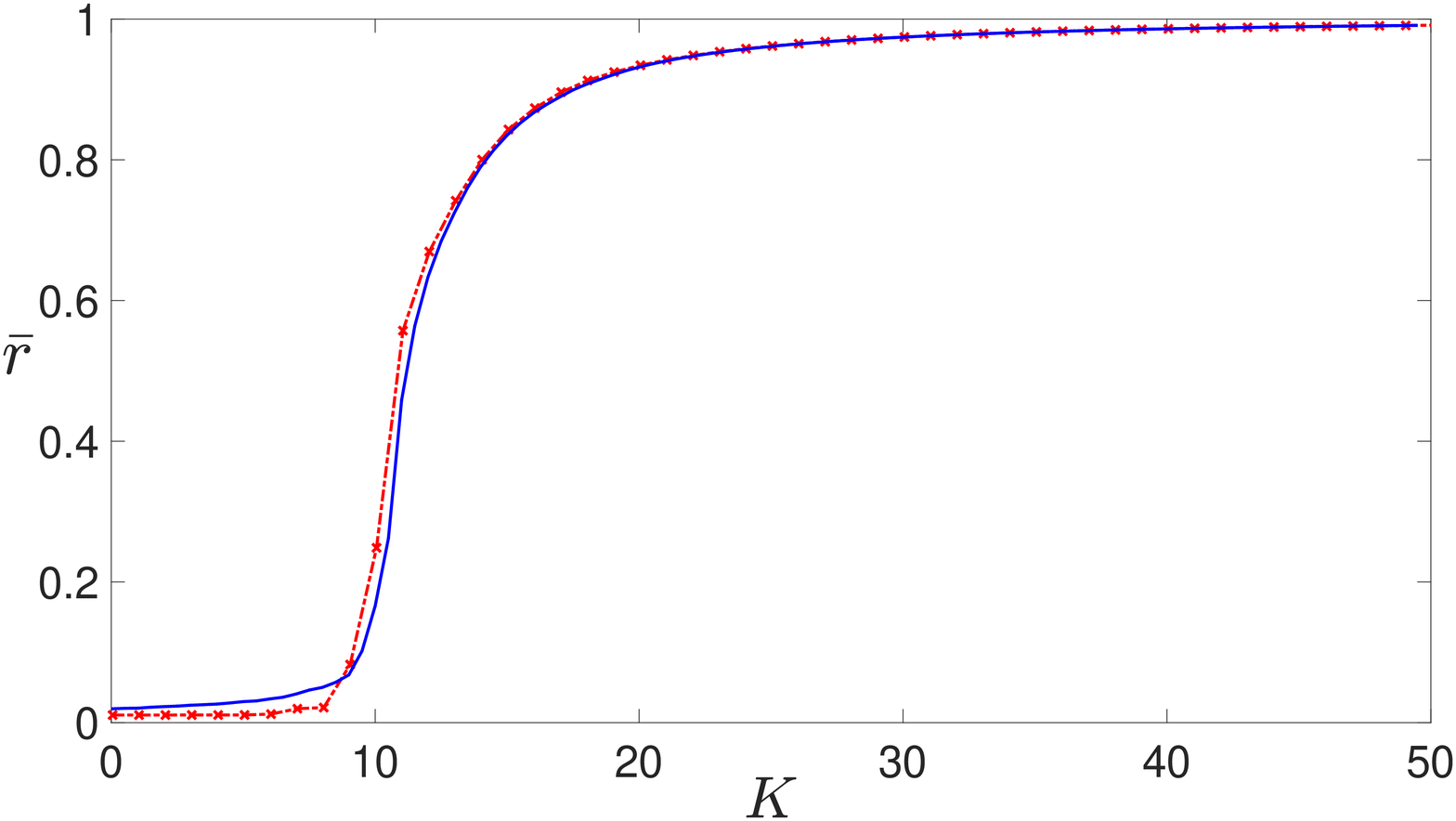}\\
	\includegraphics[width=0.5\textwidth]{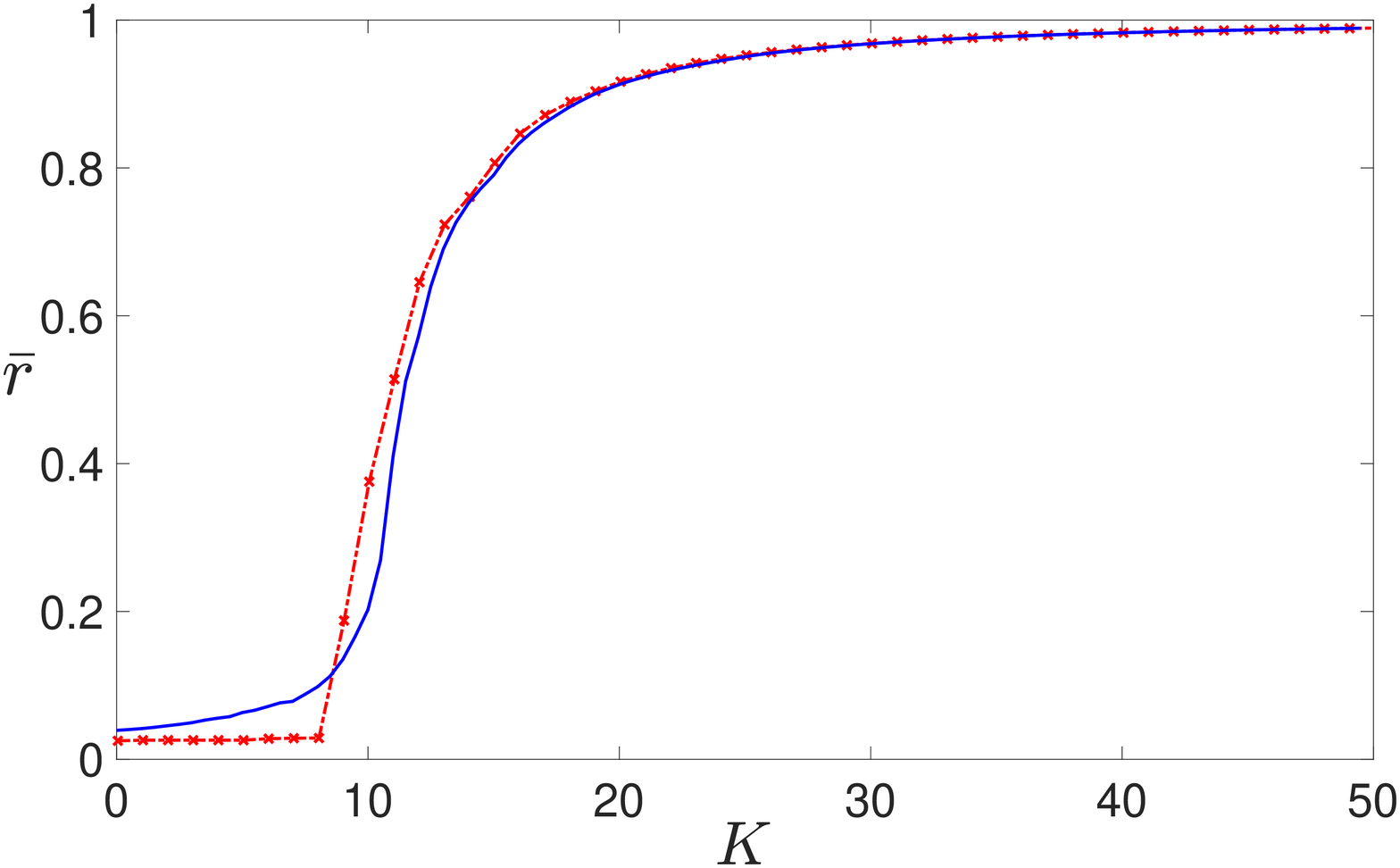}	
	\caption{Order parameter $\rbar$ as a function of the coupling strength $K$ for an ER network with normally distributed native frequencies. Depicted are results from a direct numerical integration of the Kuramoto model (\ref{e.kuramoto}) (continuous line, online blue) and from the collective coordinate approach (\ref{e.cc_gen}) using (\ref{e.rcc}) (crosses, online red). Top: ER network with $N=2000$ nodes. Bottom: ER network with $N=500$ nodes.}
	\label{fig:ERG_normal}
\end{figure}

\begin{figure}[ht]
	\centering
	\includegraphics[width=0.5\textwidth]{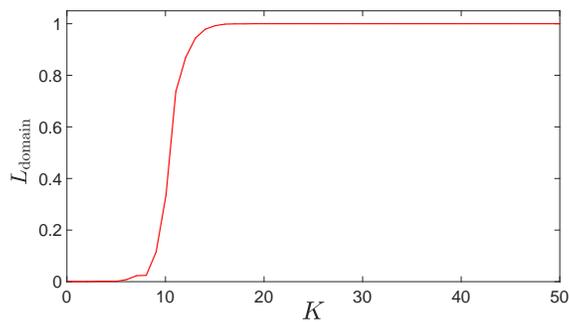}
	\caption{Normalised domain length $L_{\rm{domain}}$ as a function of the coupling strength $K$ for the ER networks depicted in Figure~\ref{fig:ERG_normal} with $N=2000$ nodes and normally distributed native frequencies.}
	\label{fig:ERG_normal_L}
\end{figure}

\change{We now present results for normally distributed frequencies. In Figure~\ref{fig:ERG_normal} it is seen that upon increasing the coupling strength $K$ the order parameter $\bar r$ becomes non-zero at some coupling strength $K_l\approx 9$ and a few oscillators with native frequencies close to the zero mean frequency locally synchronise; increasing the coupling strength allows increasingly more oscillators to synchronise, implying a continuous change of the order parameter as opposed to the hard transition in the case of uniformly distributed native frequencies seen in Figure~\ref{fig:ERG_uniform}. As for the case of the uniformly distributed native frequencies, the larger network's dynamics is very well described by the collective coordinate ansatz (\ref{e.cc_gen}) capturing both the local and the global synchronisation. The smaller network with $N=500$ nodes has a larger error describing the synchronisation behaviour accurately near the onset at $K=K_l$. This is due to, we conjecture, the presence of interacting clusters which form upon decreasing the coupling strength. In each of these smaller clusters, nodes locally synchronise and then interact. This is not described by the single-cluster ansatz (\ref{e.cc_gen}). To illustrate the local synchronisation behaviour we show in Figure~\ref{fig:ERG_normal_L} the normalised domain length
\begin{align}
L_{\rm{domain}}=\frac{N_l}{N},
\end{align}
where $N_l$ is the size of the network after discarding the unstable nodes, i.e. the size of $\C_l$, based on the linearisation matrix $L_{\rm{lin}}$ as described in Section~\ref{sec.single}. One sees clearly the smooth increase of the size of the synchronised cluster from $L_{\rm{domain}}=0$ at $K_l\approx 9$ with increasing coupling strength $K$, corresponding to the larger and larger number of oscillators joining the single synchronised cluster. At some coupling strength $K_c\approx 16$, global synchronisation sets in affecting all oscillators with $L_{\rm{domain}}=1$. We show results for the larger network with $N=2000$; the plot for the smaller network looks similar (not shown). We remark that the behaviour of ${\mathcal{F}}(\alpha)$ as described by \eqref{e.Falpha} is similar as depicted in Figure~\ref{fig:ERG_uniform_SN} for the case of uniform distributed frequencies.} \\

\begin{figure}[ht]
	\centering
	\includegraphics[width=0.5\textwidth]{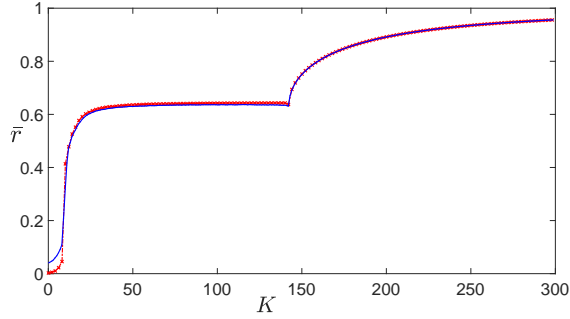}	
	\caption{Order parameter $\rbar$ as a function of the coupling strength $K$ for an ER network consisting of two coupled topological clusters with normally distributed native frequencies. The network is the same as that used for Figure~\ref{f.complexnetwork_ansatz_cluster}. Depicted are results from a direct numerical integration of the Kuramoto model (\ref{e.kuramoto}) (continuous line, online blue) and from the collective coordinate approach (\ref{e.cc_gen}) using (\ref{e.rcc}) (crosses, online red).}
	\label{fig:ERG_cluster}
\end{figure}

\begin{figure}
\begin{center}
\includegraphics[width=0.5\textwidth, height=0.25\textheight]{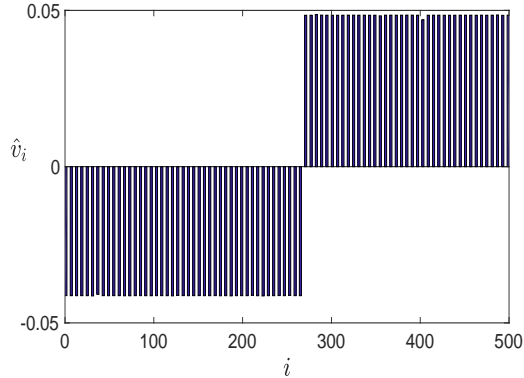}
\end{center}
\caption{\change{Normalised eigenvector $\hat v$, corresponding to the largest non-zero eigenvalue of $L_{\rm{lin}}$ for an ER network consisting of two coupled topological clusters with normally distributed native frequencies. The network is the same as that used for Figure~\ref{f.complexnetwork_ansatz_cluster} and the coupling strength is $K=142$. The eigenvector clearly separates into the two clusters with $N=270$ and $N=230$ nodes, respectively. For visual aid only every $6$th component is shown.}}
\label{f.vhatcluster}
\end{figure}

In the remainder we show that the collective coordinate ansatz (\ref{e.cc_cluster}) is able to capture the dynamics of interacting localised clusters. 
\change{Since the focus in this work is not on the identification of clusters, but rather on the ability of the collective coordinate framework to capture the dynamics of interacting clusters, we artificially generate an ER network with two well-specified topological clusters with strong intra-cluster and weak inter-cluster connectivity. In particular, we generate two ER networks, one with $N_1=270$ and one with $N_2=230$ nodes, each with a connection probability between two nodes of $p=0.05$. The two networks are now linked by $10$ randomly chosen edges to form a large network with $N=500$ nodes, in which the two networks are now tightly connected clusters. We consider again native frequencies drawn from a normal distribution $\mathcal{N}(0,0.02)$ allowing for local synchronisation within each cluster of oscillators with small absolute native frequencies. Figure~\ref{fig:ERG_cluster} shows the order parameter $\bar r$ as calculated from a long simulation of the full Kuramoto model (\ref{e.kuramoto}) as well as for the collective coordinate approach. The full simulation reveals the following synchronisation behaviour of this particular clustered network: At $K\approx 9$ the two topological clusters individually begin to locally synchronise when increasing the coupling strength from $K=0$ (cf. Figure~\ref{fig:ERG_normal}). Between $50<K\le 142$, both clusters are internally synchronised and the coupling is not strong enough to allow the two clusters to interact. In this range the order parameter is well approximated by
\begin{align*}
\rbar 
&= \frac{\Delta \omega}{2\pi}\int_0^{\frac{\Delta \omega}{2 \pi}} \big\lvert \sum_{j\in\C_1}e^{i \varphi_j} +  \sum_{j\in\C_s}e^{i \varphi_j} \big\rvert  \, dt\\
&\approx \frac{\Delta \omega}{2\pi}\int_0^{\frac{\Delta \omega}{2 \pi}}\frac{1}{N}\sqrt{N_1^2+N_2^2+2N_1N_2\cos(\Delta \omega t)}\, dt\\
&\approx 0.64,
\end{align*} 
where $\Delta \omega=0.023$ is the difference in the mean frequencies of the two respective clusters of the network under consideration. Increasing the coupling strength past $K=142$ the clusters begin to interact and increasing $K$ eventually leads to global synchronisation.\\
This path to synchronisation involving interacting clusters is remarkably well described by the collective coordinate ansatz. Starting at large values of the coupling strength the ansatz for two interacting clusters (\ref{e.cc_cluster}) with $\alpha_{1,2}$ and $f$ determined by solving (\ref{eq:red_model_m_alpha})--(\ref{eq:red_model_m_f}) accurately captures the interaction between the clusters. A snapshot of the phases for this network was already presented in Figure~\ref{f.complexnetwork_ansatz_cluster} for $K=160$. At $K\approx 142$ the collective coordinate solution becomes linearly unstable; the eigenvector $\hat v$ of the linearisation matrix $L_{\rm{lin}}$ corresponding to this instability consists of two separated parts identifying accurately the two topological clusters of the network. This is shown in Figure~\ref{f.vhatcluster} where $\hat v$ is shown for $K=142$ and the linearly unstable eigenvector clearly separates into the two clusters. For $K\le 142$ each of the two clusters is well described by the single-cluster ansatz (\ref{e.cc_gen}), each with their own independent collective coordinate $\alpha$. The stationary solutions of the evolution equation (\ref{e.alpha_c}) for the respective collective coordinates and the associated order parameter reproduce very well the collective behaviour of the full finite-size Kuramoto model.}


%


\section{Discussion and outlook}
\label{sec:out}
We \change{proposed} a collective coordinate approach for interacting Kuramoto oscillators on arbitrary networks. Our approach allows for the description of finite size networks away from the thermodynamic limit. \change{We have verified our approach against ER networks with a single synchronised cluster and with two topological clusters. Our numerical simulations on ER networks are suggestive that the collective coordinate framework is capable of quantitatively describing the collective behaviour of coupled oscillators, including the interaction of clusters. It is planned for further research to apply the collective coordinate framework to other complex network topologies such as scale-free networks and small-world network topologies where cluster formation is more prevalent. It is a highly non-trivial and, to our knowledge, an unsolved task to identify clusters for fixed coupling strength $K$. Clusters are formed in an intricate interplay between the network topology and the distribution of the native frequencies. In the numerical study of two interacting clusters, which were artificially constructed to have only a few inter-cluster links, we found that the linearisation matrix $L_{\rm{lin}}$  (which incorporates information about the network topology and the native frequencies) was able to identify the coupling strength for which clusters start to interact. Whether the linearisation matrix is able to identify clusters in more complex cases will be studied in further research. With an increasing number of clusters, the complexity of the model reduction also increases. This is particularly the case near the onset of local synchronisation where the number of partially synchronised clusters typically grows and near onset of synchronisation the collective coordinate approach might be computationally as costly as simulating the full system. The computational cost, however, if only a few clusters $M\ll N$ are present is significantly lower than simulating the whole system. When there is only a single cluster present no dynamics needs to be evolved and the problem reduces to finding roots (cf (\ref{e.Falpha})). For $M$ clusters one needs to evolve $2(M-1)$ equations for the collective coordinates $\alpha_m$ and $f_m$. In our experience the length of the numerical simulation to obtain converged behaviour for the order parameter is much less than for the full $N$-dimensional Kuramoto model. Whereas the Kuramoto model may be a highly stiff dynamical system depending on the native frequencies used, the evolution equations for the collective coordinates are not stiff. Hence simulating the collective coordinate system allows for a larger time step to be used in the numerical simulations.}

\change{Our simulations did not consider the important case of possible spatial correlations of the native frequencies. When the native frequencies are positively (negatively) correlated among neighbouring nodes, synchronisation is favoured (inhibited). Determining whether the collective coordinate framework can reproduce this complex interplay between topology and dynamics is planned for further research.
}


\section*{Acknowledgements}
GAG acknowledges support from the Australian Research Council, grant DP180101991. EJH gratefully acknowledges the donation from Judith and David Coffey.


\bibliographystyle{plain}

\begin{thebibliography}{10}

\bibitem{AcebronEtAl05}
Juan~A. Acebr\'on, L.~L. Bonilla, Conrad~J. P\'erez~Vicente, F\'elix Ritort,
  and Renato Spigler.
\newblock The {K}uramoto model: {A} simple paradigm for synchronization
  phenomena.
\newblock {\em Rev. Mod. Phys.}, 77:137--185, Apr 2005.

\bibitem{ArenasEtAl08}
Alex Arenas, Albert Diaz-Guilera, J\"urgen Kurths, Yamir Moreno, and Changsong
  Zhou.
\newblock Synchronization in complex networks.
\newblock {\em Phys. Rep.}, 469(3):93--153, 2008.

\bibitem{BhowmikShanahan12}
D.~Bhowmik and M.~Shanahan.
\newblock How well do oscillator models capture the behaviour of biological
  neurons?
\newblock In {\em The 2012 International Joint Conference on Neural Networks
  (IJCNN)}, pages 1--8, 2012.

\bibitem{BredeKalloniatis16}
Markus Brede and Alexander~C. Kalloniatis.
\newblock Frustration tuning and perfect phase synchronization in the
  {K}uramoto-{S}akaguchi model.
\newblock {\em Phys. Rev. E}, 93:062315, Jun 2016.

\bibitem{Crawford94}
John~David Crawford.
\newblock Amplitude expansions for instabilities in populations of
  globally-coupled oscillators.
\newblock {\em Journal of Statistical Physics}, 74(5-6):1047--1084, 1994.

\bibitem{DoerflerBullo14}
Florian D{\"o}rfler and Francesco Bullo.
\newblock Synchronization in complex networks of phase oscillators: {A} survey.
\newblock {\em Automatica}, 50(6):1539 -- 1564, 2014.

\bibitem{FilatrellaEtAl08}
G.~Filatrella, A.~H. Nielsen, and N.~F. Pedersen.
\newblock Analysis of a power grid using a {K}uramoto-like model.
\newblock {\em The European Physical Journal B}, 61(4):485--491, Feb 2008.

\bibitem{Golub}
G.~H. Golub and Ch.~F.~Van Loan.
\newblock {\em {Matrix Computations}}.
\newblock The Johns Hopkins University Press, Baltimore, 3rd edition, 1996.

\bibitem{Gottwald15}
Georg~A. Gottwald.
\newblock Model reduction for networks of coupled oscillators.
\newblock {\em Chaos}, 25(5):053111, 12, 2015.

\bibitem{Gottwald17}
Georg~A. Gottwald.
\newblock Finite-size effects in a stochastic {K}uramoto model.
\newblock {\em Chaos: An Interdisciplinary Journal of Nonlinear Science},
  27(10):101103, 2017.

\bibitem{Kuramoto}
Y.~Kuramoto.
\newblock {\em Chemical {O}scillations, {W}aves, and {T}urbulence}, volume~19
  of {\em Springer Series in Synergetics}.
\newblock Springer-Verlag, Berlin, 1984.

\bibitem{MartensEtAl09}
E.~A. Martens, E.~Barreto, S.~H. Strogatz, E.~Ott, P.~So, and T.~M. Antonsen.
\newblock Exact results for the {K}uramoto model with a bimodal frequency
  distribution.
\newblock {\em Phys. Rev. E}, 79:026204, Feb 2009.

\bibitem{MarvelEtAl09}
Seth~A. Marvel, Renato~E. Mirollo, and Steven~H. Strogatz.
\newblock Identical phase oscillators with global sinusoidal coupling evolve by
  {M}\"obius group action.
\newblock {\em Chaos}, 19(4):043104, 11, 2009.

\bibitem{Note1}
For stochastic Kuramoto models, the mean phase experiences non-trivial
  diffusive behaviour which can also be captured by the collective coordinate
  framework, see \cite {Gottwald17}.

\bibitem{Note2}
Note that $L$ has a single zero eigenvalue with corresponding eigenvector $V_1$
  satisfying $V_1^T\omega =0$, and $N-1$ repeated eigenvalues $\lambda =N$.
  Using an eigenvalue decomposition, write $L^+\omega =V D^+V^T\omega =\lambda
  ^{-1} \omega $ which implies (\ref {e.cc_alltoall}).

\bibitem{Note3}
See \cite {Gottwald15} for an example of two interacting clusters in an
  all-to-all coupling network with a bimodal frequency distribution.

\bibitem{OsipovEtAl}
Grigory~V. Osipov, J{\"u}rgen Kurths, and Changsong Zhou.
\newblock {\em Synchronization in {O}scillatory {N}etworks}.
\newblock Springer Series in Synergetics. Springer, Berlin, 2007.

\bibitem{OttAntonson08}
Edward Ott and Thomas~M. Antonsen.
\newblock Low dimensional behavior of large systems of globally coupled
  oscillators.
\newblock {\em Chaos}, 18(3):037113, 6, 2008.

\bibitem{PikovskyEtAl}
A~Pikovsky, M.~Rosenblum, and J.~Kurths.
\newblock {\em {Synchronization: {A} {U}niversal {C}oncept in {N}onlinear
  {S}ciences}}.
\newblock Cambridge University Press, Cambridge, 2001.

\bibitem{PikovskyRosenblum08}
Arkady Pikovsky and Michael Rosenblum.
\newblock Partially integrable dynamics of hierarchical populations of coupled
  oscillators.
\newblock {\em Phys. Rev. Lett.}, 101:264103, Dec 2008.

\bibitem{PikovskyRosenblum11}
Arkady Pikovsky and Michael Rosenblum.
\newblock Dynamics of heterogeneous oscillator ensembles in terms of collective
  variables.
\newblock {\em Physica D}, 240(9--10):872 -- 881, 2011.

\bibitem{PintoSaa15}
Rafael~S. Pinto and Alberto Saa.
\newblock Optimal synchronization of {K}uramoto oscillators: a dimensional
  reduction approach.
\newblock {\em Phys. Rev. E (3)}, 92(6):062801, 6, 2015.

\bibitem{RodriguesEtAl16}
Francisco~A. Rodrigues, Thomas K.~DM. Peron, Peng Ji, and J{\"u}rgen Kurths.
\newblock The {K}uramoto model in complex networks.
\newblock {\em Physics Reports}, 610:1 -- 98, 2016.

\bibitem{SheebaEtAl08}
Jane~H. Sheeba, Aneta Stefanovska, and Peter V.~E. McClintock.
\newblock Neuronal synchrony during anesthesia: {A} thalamocortical model.
\newblock {\em Biophysical Journal}, 95(6):2722--2727, 2008.

\bibitem{Strogatz00}
Steven~H. Strogatz.
\newblock From {K}uramoto to {C}rawford: {E}xploring the onset of
  synchronization in populations of coupled oscillators.
\newblock {\em Physica D}, 143(1-4):1--20, 2000.

\bibitem{WattsStrogatz98}
Duncan~J. Watts and Steven~H. Strogatz.
\newblock Collective dynamics of `small-world' networks.
\newblock {\em Nature}, 393:440 EP --, 06 1998.

\end{thebibliography}

\end{document}